\documentclass[aps,prd,groupedaddress,onecolumn,floatfix,nofootinbib,showpacs,showkeys]{revtex4-1}

\usepackage{graphicx}% Include figure files
\usepackage{dcolumn}% Align table columns on decimal point
\usepackage{bm}% bold math
\usepackage{amsmath}
\usepackage{epstopdf}
\usepackage[flushleft]{threeparttable}

\usepackage[thinlines]{easytable}
\usepackage{array}
\usepackage{booktabs}
\usepackage{multirow}

\begin{document}

\title{Properties of Rotating Neutron Star in Density-dependent Relativistic Mean-field Models}
\author{R.~Riahi}
\email{r.riahi@iaushk.ac.ir}
\author{S.~ Z.~Kalantari}
\email{zafar@cc.iut.ac.ir}
\affiliation{Department of Physics, Isfahan University of Technology, 84156-83111, Iran}

\date{\today}

\begin{abstract}
Equilibrium sequences were developed for rotating neutron stars in the relativistic mean-field interaction framework using four density-dependent equations of state(EOS) for the neutron star matter. These sequences were constructed for the observed rotation frequencies of 25, 317, 346, 716, and 1122 Hz. The bounds of sequences, the secular axisymmetric instability, static, and Keplerian sequences were calculated in each model to determine the stability region. The gravitational mass, quadrupole moment, polar, forward and backward redshifts, and Kerr parameter were calculated according to this stability region, and the allowable range of these quantities were then determined for each model. According to the results, DDF and DD-ME$\delta$ were unable to properly describe the low-frequency neutron stars, PSR J0348+432, PSR J1614-2230 , and PSR J0740+6620 rotate at a frequency of 25, 317, and 346 Hz, respectively. On the other hand, all the selected EOSs properly described the rotation of PSR J1748-244ad, and PSR J1739-285 at a frequency of 716 and 1122 Hz, respectively. The mass of these stars was, therefore, in the range of [0.68, 2.14]M$_{\odot}$ and [1.67, 2.24]M$_{\odot}$, respectively.
The polar, forward and backward redshifts, and the quadrupole moment were calculated in all selected rotating frequencies and the Keplerian sequence. The results were consistent with observations. Confirming the mass of $1.5^{+0.4}_{-1.0}M_{\odot}$ for EXO 0748-676, our result, $Z_{P}\approx 0.3$, will be close to the observed value, and the EOSs used in this study properly describe this star.   
Interestingly, the extremum of Kerr parameter, polar, forward and backward redshifts in all models reached constant values of, a/M$\approx$ 0.7, Z$_{p}\approx$ 0.8, Z$_{eq}^{f}\approx$ -0.3 and Z$_{eq}^{b}\approx$ 2.2, respectively. These behaviors of redshifts and Kerr parameter are approximately independent of EOS. The observed behaviors must evaluate by other EOSs to find universal relations for these quantities. Also, a limit value was found for each of these parameters. In this case where these parameters are greater than the limit value, the star can rotate at a frequency equal to or greater than $\nu$= 1122 Hz. 
\end{abstract}



\maketitle


%%%%%%%%%%%%%%%%%%%%%%%%%%%%%%%%%%%%%%%%%%%%%%%%%%%%%%%%
%%%%%%%%%%%%%%%%%%%%%%%%%%%%%%%%%%%%%%%%%%%%%%%%%%%%%%%%
\section{Introduction}
%%%%%%%%%%%%%%%%%%%%%%%%%%%%%%%%%%%%%%%%%%%%%%%%%%%%%%%%
%%%%%%%%%%%%%%%%%%%%%%%%%%%%%%%%%%%%%%%%%%%%%%%%%%%%%%%%
Numerous studies have been conducted on the properties of static neutron stars\cite{Lattimer2000,LI2018234,Li_2019,PhysRevC.100.015809,PhysRevD.101.034017}. The rotating neutron stars have not been extensively studied despite discovering the first millisecond pulsar in 1982\cite{Backer1982}. Because solving Einstein's field equations for rotating object is a very complicated task, the rotating stars must be modeled numerically. In a pioneering work, Bonazzola and Schnider\cite{1974ApJ...191..273B}, and Butterworth\cite{1976ApJ...204..561B} considered incompressible fluids and polytropic EOS for modeling rotating stars. Friedman \textit{et al.}\cite{1986ApJ...304..115F} used some realistic EOSs for investigating neutron star matter in a general relativity framework. Cook \textit{et al.}\cite{1994ApJ...424..823C} studied  the quasi-stationary evolution of isolated neutron stars based on a formalism introduced by Komatsu \textit{et al.}\cite{1989MNRAS.237..355K} (KEH formalism). Also, Datta \textit{et al.}\cite{1998A&A...334..943D} used the KEH formalism to report the equilibrium sequences of rotating neutron stars using realistic EOS. The equilibrium sequences of rotating stars are essential to model and investigate some astrophysical phenomena such as millisecond pulsars, Quasi Periodic Oscillations (QPOs), and low-mass X-ray binaries (LMXBs). Some quantities such as mass, radius, quadrupole moment, and the Keplerian frequency of a particle orbiting around a rotating star can be determined from these sequences. These quantities play a key role in studying LMXBs and QPOs emission.
Salgado \textit{et al.}\cite{1994A&A...291..155S,1994A&AS..108..455S} applied the spectral-based approach developed by Bonazzola \textit{et al.}\cite{1993A&A...278..421B}  to a variety of EOSs of the neutron star matter. Laarakkers \textit{et al.}\cite{0004-637X-512-1-282} used the public code, RNS, developed based on the KEH formalism by Stergioulas and Friedman, to calculate the quadrupole moment of rotating neutron stars. Cipolletta \textit{et al.}\cite{PhysRevD.92.023007,PhysRevD.96.024046} presented the equilibrium sequences of rotating neutron star, the binding energy and angular momentum of a test particle at the last stable circular orbit (LSO) on the equatorial plane of the neutron star with the help of RNS. Li \textit{et al.} generated a set of EOSs with covariant density functional theory and investigated the properties of static and fast rotating compact stars with and without $\delta$-resonance-admixed hyperonic core compositions\cite{LI2020135812}. Sedrakian et al. investigated the possibility that the light companion in the binary compact object coalescence event GW190814 is a hyper nuclear star. They used covariant density functional theory of hypernuclear matter and considered both static configurations and fast rotating configurations at its Keplerian frequency\cite{PhysRevD.102.041301}. Using RNS with EOSs on the causality surface and satisfying all known constraints from both nuclear physics and astrophysics, Zhang and Li investigated the GW190814's component, and concluded it can be a super-fast pulsar spinning faster than 971 Hz\cite{Zhang_2020}.

A proper EOS for a high-density matter is the starting point for investigating the structure of neutron stars. However, there exists much theoretical discussion on proper EOS\cite{PhysRevC.98.035804,arXiv:2007.03799}, and each EOS determines different equilibrium sequences of static and rotating stars. 
Despite numerous studies on high-density matter, there is still no agreement on its composition and EOS, especially for densities several times greater than the normal nuclear matter density. The core of a neutron star is considered a nuclear matter in the beta equilibrium, which is an electrically neutral fluid composed of neutron, proton, electron and muon. Some additional degrees of freedom, hyprons, Kaon or a deconfined phase of quark matter can also be imposed. The possible appearance of such an exotic core significantly affects the properties of neutron stars\cite{LI2018234,Li_2019,PhysRevC.100.015809,PhysRevD.101.034017,Haensel_2016,1994ApJ...423..659B,1996A&A...305..871B,2017IJMPD..2630004B,2016EPJA...52...58B}. In this study, the neutron star core is assumed to be composed of only neutron, proton, electron and muon as a beta stable matter.

The relativistic mean field (RMF) approach with density-dependent meson-nucleon couplings, includes the isoscalar scalar channel, $\sigma$ meson, the isoscalar vector channel, $\omega$ meson, and the isovector vector channel, $\rho$ meson, is widely used to describe the neutron stars. However, ignoring the isovector scalar channel seems to be illogical at high densities in strongly isospin, asymmetric matter such as neutron stars\cite{1997PhLB..399..191K,PhysRevC.67.015203}. Phenomenological studies\cite{PhysRevC.70.015203,PhysRep.410.335,Int.J.Mod.PhysE.9.1815,PhysRevC.80.045808,PhysRevC.80.025806,Res. Astron.Astrophys.10.1255} and microscopic investigations\cite{Nucl.Phys.A596.684,Int.J.Mod.Phys.E.7.301} revealed that mean-field models neglecting the $\delta$ meson fail to take into account important ingredients in describing a highly asymmetric nuclear matter, especially at high densities. The splitting of the effective mass of proton and neutron can affect transport properties in heavy-ion reactions and neutron stars; also, the proton fraction of $\beta$-stable matter in neutron stars can increase.
The $\delta$ meson is introduced in RMF\cite{PhysRevC.70.058801,Gaitanos200424}, and the best tuning was performed based on the selected microscopic calculations presented by Roca-Maza \textit{et al.}\cite{PhysRevC.84.054309}.

The properties of nuclei can be described successfully in the mean-field interactions framework. This approach describes the nuclear many-body problem as an energy density functional theory(DFT). Therefore, the construction of such a universal theory is considered a major goals of nuclear physics\cite{PhysRevC.71.024312}. In an early work, Boguta and Bodmer\cite{1977NuPhA.292..413B} used a phenomenological nonlinear meson interaction to present a phenomenological density-dependent functional theory. The nonlinear models were then changed to explicit density dependent models. The latter produced remarkable results and successfully described asymmetric nuclear matter, neutron matter and nuclei far from stability. Accordingly, the relativistic mean-field interactions with density-dependent meson-nucleon couplings, DDH$\delta$, TW99, DD-F and DD-ME$\delta$, were used in this study to describe the rotating neutron stars.

Measuring the redshift of spectral lines produced in the neutron star photosphere is considered a useful method for determining the fundamental properties of neutron stars. The redshift measurement of a neutron star provides a direct constraint on the mass-to-radius ratio. On the other hand, a fast-rotating neutron star is not spherically symmetric, because the rotation makes a deformation in the distribution of the stellar mass. Subsequently, this oblateness creates a distortion in the gravitational field outside the star, which is measured by the quadrupole moment tensor. In this work, the redshift and quadrupole moment of the rotating neutron stars are calculated at the observed rotating frequencies.  

The equilibrium sequences of rotating neutron stars are constructed in the general relativity for the selected EOSs. These sequences are calculated at a rotation frequency of 25 Hz related to PSR J0348+432 with a mass of (2.01$\pm$0.04)M$_{\odot}$\cite{Antoniadis1233232}, 317 Hz related to PSR J1614-2230 with a mass of (1.97$\pm$0.04)M$_{\odot}$\cite{Demorest2010/10/28/print} and, 346 Hz related to PSR J0740+6620 with a mass of $2.14^{+0.1}_{-0.09} M_{\odot}$ \cite{cromartie2019relativistic}, as the most massive rotating stars yet found, and at rotation frequency 716 Hz for PSR J1748-2446ad\cite{2006Sci...311.1901H} and 1122 Hz for X-ray transient XTE J1739-285\cite{1538-4357-657-2-L97}, as the most rapidly rotating neutron stars observed so far. An oscillation at 1122 Hz was announced by Kaaret \textit{et. al.}\cite{1538-4357-657-2-L97} but has not been yet confirmed. This can be related to different choices of time windows for the 4-s FFTs\cite{7c6fd5014fc848d89f7d9b5f9cbd2da9}. Using the observed data, we want to investigate which EOS is appropriate to describe these stars and which one rules out. The unmeasured redshift and quadrupole moment that may be measured in the future are also calculated. The behavior of these quantities is also investigated to find the behavior independent of EOS to establish universal relations\cite{PhysRevD.99.043004}.
%%%%%%%%%%%%%%%%%%%%%%%%%%%%%%%%%%%%%%%%%%%%%%%%%%%%%%%%
%%%%%%%%%%%%%%%%%%%%%%%%%%%%%%%%%%%%%%%%%%%%%%%%%%%%%%%%
\section{Equation of State}
\begin{figure} 
	\includegraphics[width=8.5cm]{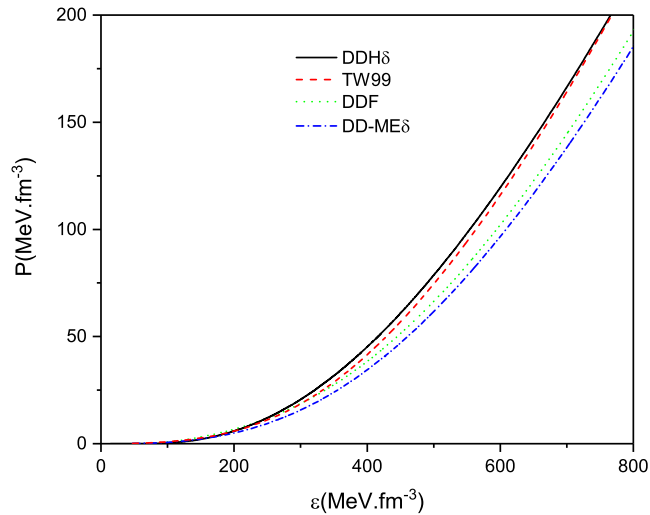}
		\caption{(Color online) The pressure of neutron star matter as a function of the energy density for each EOS. }\label{EOS}
\end{figure}

Equation of state (EOS), relation between energy density and pressure, is an essential requirement for describing the macroscopic properties of stars. The EOS is used as an input to the Einstein's field equations. Non-relativistic\cite{PhysRevC.85.035201} and relativistic models\cite{PhysRevC.90.055203} have been used to obtain stellar properties. Herein, the density-dependent relativistic models are used in the mean field approximation. DDH$\delta$\cite{Gaitanos200424}, TW99 \cite{Typel1999}, DD-F \cite{PhysRevC.71.064301}, and DD-ME$\delta$\cite{PhysRevC.84.054309} are used to describe the core of the star. Figure \ref{EOS} shows the pressure of neutron star matter as a function of the energy density for each EOS. The Baym-Pethick-Sutherland (BPS) equation of state\cite{1971ApJ...170..299B} to describe the crust of the star.

%%%%%%%%%%%%%%%%%%%%%%%%%%%%%%%%%%%%%%%%%%%%%%%%%%%%%%%%

%%%%%%%%%%%%%%%%%%%%%%%%%%%%%%%%%%%%%%%%%%%%%%%%%%%%%%%%
%%%%%%%%%%%%%%%%%%%%%%%%%%%%%%%%%%%%%%%%%%%%%%%%%%%%%%%%
%%%%%%%%%%%%%%%%%%%%%%%%%%%%%%%%%%%%%%%%%%%%%%%%%%%%%%%%
\section{Numerical Solution of Einstein's field equations}
%%%%%%%%%%%%%%%%%%%%%%%%%%%%%%%%%%%%%%%%%%%%%%%%%%%%%%%%
%%%%%%%%%%%%%%%%%%%%%%%%%%%%%%%%%%%%%%%%%%%%%%%%%%%%%%%%
Many numerical methods have been developed to investigate the structure and gravitational field of relativistic, axisymmetric, stationary and uniform rotating stars since the 1970s \cite{1972ApJ...176..195W,1976ApJ...204..561B,1974ApJ...191..273B,1986ApJ...304..115F,PhysRevLett.62.3015,1989MNRAS.237..355K,1989MNRAS.239..153K,1990ApJ...355..241L,1992LNP...410..305N,1995ApJ...444..306S}.
In this study, the recently developed multi-domain spectral method (AKM-method)\cite{AnsorgM.2003} is used. In AKM-method, the entire spatial domain is divided into several sub-domains, and all physical field quantities are expressed in a spectral expansion with respect to coordinates. When one of the selected domain boundaries coincides the star's surface, the representations of field quantities are smooth, then this leads to a very precise approximation of the desired quantities. In this manner, it is possible to circumvent the Gibbs phenomenon on the star's surface. This happens when non-smooth physical fields (\textit{e.g.} the mass-energy density) are expressed in terms of a spectral expansion and affects the spectral convergence rate.
In this study, the library LORENE (http://www.lorene.obspm.fr) is used for the numerical solution of the Einstein's field equations.

%%%%%%%%%%%%%%%%%%%%%%%%%%%%%%%%%%%%%%%%%%%%%%%%%%%%%%%%
%%%%%%%%%%%%%%%%%%%%%%%%%%%%%%%%%%%%%%%%%%%%%%%%%%%%%%%%
\section{Results}
%%%%%%%%%%%%%%%%%%%%%%%%%%%%%%%%%%%%%%%%%%%%%%%%%%%%%%%%
%%%%%%%%%%%%%%%%%%%%%%%%%%%%%%%%%%%%%%%%%%%%%%%%%%%%%%%%
\subsection{Equilibrium Configurations}
In order to find an equilibrium sequence of states of the rotating neutron star in each model, a constant $\Omega$ and a variable central energy density in a certain range were used. The radius-mass sequence at a constant $\Omega$ or J is the most important representation of a neutron star. The sequences are restricted to the following three limits. I) the first limit on the stability of rotating stars is the static configurations in which $\Omega=0$. For each mass, it indicates the minimum equatorial radius of the star. With slightly more compaction, the star becomes unstable and collapses to a black hole. For the static sequence, the first turning point of this with increasing central energy density, $\epsilon_{c}$, corresponds to the maximum stable mass of star. This point can be found by $\frac{\partial M}{\partial\epsilon_{c}}=0$. II) The second limit is the Keplerian or mass-shedding sequence. A particle is in a stable circular orbit at the equator of a star when the centrifugal and gravitational forces are in balance. Considering this condition, the Kepler angular velocity, referring to briefly as the Kepler frequency, is found. At the equator of star, fluid elements are dislodged at velocity greater than this. III) The third limit, at which instability to quasi-radial mode sets in, is called secular axisymmetric instability. If the mass in a constant angular momentum is plotted against the central energy density $\epsilon_{c}$, the mass reaches a peak. This indicates the secular axisymmetric instability and sets in at\cite{1988ApJ...325..722F}:
\begin{equation}
\left.\frac{\partial M(\epsilon_{c},J)}{\partial\epsilon_{c}}\right\vert_{J=constant}=0,
\end{equation}  
where J is the angular momentum. The intersection of this curve with the Keplerian sequence gives the fastest possible configuration allowed by the EOS. Table \ref{table:fastest} lists the properties of the fastest rotating neutron stars can be described by the selected EOSs. One can conclude that the rotation frequency of the fastest observed star, 1122 Hz, does not impose a strong constraint on the EOS(In the first column, the EOS are listed in order of decreasing stiffness). By using the above-mentioned limits, the range of physical quantities such as mass, radius, redshift, and quadruple moment of the rotating neutron stars can be found.

%%%%%%%%%%%%%%%%%%%%%%%%%%%%%%%%%%%%%%%%%%%%%%%%%%%%%%%%%%%%%%%%%%%%%%%%% 
\begin{table*}
	\caption{Properties of the fastest rotating neutron star for each model}
	{	\begin{tabular}{@{}cccccccccc@{}} \hline\hline\\
			%\begin{tabular*}{p{0.25\linewidth}p{0.25\linewidth}p{0.25\linewidth}p{0.25\linewidth}}
			
			%		\multirow{2}{*}{Model} &\multicolumn{8}{c}{Fastest Star} \\[1ex]
			%		\cline{2-9} \\[-1ex]
			Model &   M(M$_{\odot}$)  & R$_{eq}$(km)& $\nu$(Hz) & (a/M)$_{max}$ & Z$_{p}$ ~ & Z$^{f}_{eq}$ ~ & Z$^{b}_{eq}$ & Q($10^{43}$ gr.cm$^2$)\\[1ex]
			\hline\\[1ex]
			DDH$\delta$ & 2.50 & 14.77 & 1594.24 & 0.70 & 0.75 ~ & -0.33 ~ & 2.17 & -45.37 \\[1ex]
			TW99 &  2.46 & 14.12 & 1673.38 & 0.69 & 0.78 ~ & -0.34 ~ & 2.28 & -43.50 \\[1ex]
			DD-F & 2.36 & 13.71 & 1718.33 &  0.70 & 0.77 ~ & -0.34 ~ & 2.24 & -40.30 \\[1ex]
			DD-ME$\delta$ & 2.26 & 13.49 & 1768.77 & 0.70 & 0.76 ~ & -0.33 ~ & 2.19 & -32.94 \\[1ex]
			\botrule
		\end{tabular}\label{table:fastest}}
\end{table*}
%%%%%%%%%%%%%%%%%%%%%%%%%%%%%%%%%%%%%%%%%%%%%%%%%%%%%%%%%%%%%%%%%%%%%%%%% 
The equilibrium sequences were constructed at a frequency of 25, 317, 346, 716, and 1122 Hz. The range of physical quantities such as mass, radius, redshift and quadrupole moment of stars can be obtained from these calculations for each model. 
The parameters of the rotating neutron stars extracted for each selected EOS are presented in Tables \ref{table:DDHd:rotation}-\ref{table:DD-ME:rotation} (In each column the upper and lower values in each row represent the parameter of star corresponding to maximum and minimum mass, respectively). At low frequencies, the effect of rotation on the star mass is small and the rotating star mass is equal to the static star mass, approximately. At the rotating frequencies of $\nu<350$ Hz, the maximum calculated masses are 1.97, and 1.94$M_{\odot}$, approximately, by the DD-F and DD-ME$\delta$. On the other hand, the maximum observed masses at the rotating frequencies of 25, 317, and 346 Hz are 2.05, 2.01, and 2.24 $M_{\odot}$, respectively. Therefore, these models are so soft to sustain the maximum observed masses at the rotating frequencies of 25, 317, and 346 Hz. These EOSs are, therefore, ruled out in describing these rotating neutron stars.
For describing J1748-2446ad and J1739-285 rotating at 716 and 1122 Hz, respectively, all selected EOSs seem appropriate, and the calculated ranges of mass are [0.68 -2.14]M$_{\odot}$ and [1.67 ,2.24] M$_{\odot}$, respectively.

%%%%%%%%%%%%%%%%%%%%%%%%%%%%%%%%%%%%%%%%%%%%%%%%%%%%%%%%%%%%%%%%%%%%%
\begin{table}[ph]
	\caption{Properties of the rotating neutron star in the observed frequencies for DDH$\delta$ model.} 
	{\begin{tabular}{@{}cccccc@{}} \hline\hline\\[1ex]
			
			$\nu$(Hz) & M(M$_{\odot}$)  & Z$_{p}$ & ~Z$_{eq}^{f}$ & ~Z$_{eq}^{b}$ & Q($10^{43}$gr.cm$^{2}$) \\ [1ex]
			\hline\\[1ex]
			&  & 0.44 & ~ 0.43 & ~0.45 & $-0.05$  \\[-1ex]
			\raisebox{1.5ex}{25} & \raisebox{1.5ex}{2.01$\pm$0.04\cite{Antoniadis1233232}}
			&  0.38 & ~ 0.37 & ~0.40 & $-0.09$  \\[1ex]
			&  & 0.40 & ~ 0.25 & ~0.57 & $-2.00$  \\[-1ex]
			\raisebox{1.5ex}{317} & \raisebox{1.5ex}{1.97$\pm$0.04\cite{Demorest2010/10/28/print}}
			&   0.37 & ~ 0.22 & ~0.52& $-2.37$  \\[1ex]
			& & No Stars   &  &  &   \\[-1ex]
			\raisebox{1.5ex}{346} & \raisebox{1.5ex}{2.14$^{+0.1}_{-0.09}$\cite{cromartie2019relativistic}}
			&  0.43 & ~ 0.26 & ~0.62 & $-1.72$  \\[1ex]
			& 2.14 & 0.53 & ~ 0.17 & ~0.96 & $-5.39$  \\[-1ex]
			\raisebox{1.5ex}{716} & 
			0.96 & 0.14 & $-0.20$ & ~0.47 & $-22.02$  \\[1ex]
			& 2.24 & 0.56 & $-0.05$ & ~1.31& $-19.39$  \\[-1ex]
			\raisebox{1.5ex}{1122} &
			2.04  & 0.40 & $-0.28$ & ~1.16 & $-52.28$  \\[1ex]
			
			\botrule
		\end{tabular}
		\label{table:DDHd:rotation}}
\end{table}
%%%%%%%%%%%%%%%%%%%%%%%%%%%%%%%%%%%%%%%%%%%%%%%%%%%%%%%%%%%%%%%%%%%%%%%%%%
%%%%%%%%%%%%%%%%%%%%%%%%%%%%%%%%%%%%%%%%%%%%%%%%%%%%%%%%%%%%%%%%%%%%%
\begin{table}[ph]
	\caption{Properties of the rotating neutron star in the observed frequencies for TW99 model.}  
	{\begin{tabular}{@{}cccccc@{}} \hline\hline\\[1ex]
			$\nu$(Hz) & M(M$_{\odot}$)   & Z$_{p}$ & ~ Z$_{eq}^{f}$ & Z$_{eq}^{b}$ & Q($10^{43}$gr.cm$^{2}$) \\ [1ex]
			\hline\\[1ex]
			&  & 0.49 & ~ 0.48 & ~0.51 & $-7.5\times 10^{-3}$  \\[-1ex]
			\raisebox{1.5ex}{25} & \raisebox{1.5ex}{2.01$\pm$0.04}
			&  0.42 & ~ 0.41 & ~0.44 & $-8.24\times 10^{-3}$  \\[1ex]
			& & 0.45 & ~ 0.30 & ~0.61 & $-1.14$  \\[-1ex]
			\raisebox{1.5ex}{317} & \raisebox{1.5ex}{1.97$\pm$0.04}
			& 0.40 & ~ 0.25 & ~0.55 & $-1.34$  \\[1ex]
			& & No Stars   &  &  &   \\[-1ex]
			\raisebox{1.5ex}{346} & \raisebox{1.5ex}{2.14$^{+0.1}_{-0.09}$}
			&  0.49 & ~ 0.31 & ~0.67 & $-1.21$  \\[1ex]
			&2.10 &  0.56 & ~ 0.19 & ~0.98 & $-4.47$  \\[-1ex]
			\raisebox{1.5ex}{716} & 
			0.82  & 0.12 & $-0.19$ & ~0.44 & $-13.74$  \\[1ex]
			& 2.18  & 0.59 & $\approx$ 0 & ~1.33 & $-13.54$  \\[-1ex]
			\raisebox{1.5ex}{1122} &
			1.8  & 0.35 & ~ 0.27 & ~1.04 & $-40.57$  \\[1ex]
			
			\botrule
		\end{tabular}
		\label{table:TW99:rotation}}
\end{table}
%%%%%%%%%%%%%%%%%%%%%%%%%%%%%%%%%%%%%%%%%%%%%%%%%%%%%%%%%%%%%%%%%%%%%%%%%%
%%%%%%%%%%%%%%%%%%%%%%%%%%%%%%%%%%%%%%%%%%%%%%%%%%%%%%%%%%%%%%%%%%%%%
\begin{table}[ph]
	\caption{Properties of the rotating neutron star in the observed frequencies for DD-F model} 
	{\begin{tabular}{@{}cccccc@{}} \hline\hline\\[1ex]
			$\nu$(Hz) & M(M$_{\odot}$)  & Z$_{p}$ & ~ Z$_{eq}^{f}$ & Z$_{eq}^{b}$ & Q($10^{43}$gr.cm$^{2}$) \\ [1ex]
			\hline\\[1ex]
			& & No Stars   &  & & \\[-1ex]
			\raisebox{1.5ex}{25} & \raisebox{1.5ex}{2.01$\pm$0.04\cite{Antoniadis1233232}}
			& No Stars  &  & & \\[1ex] 
			& & No Stars   & & &  \\[-1ex]
			\raisebox{1.5ex}{317} & \raisebox{1.5ex}{1.97$\pm$0.04\cite{Demorest2010/10/28/print}}
			&  0.45 & ~ 0.30 & 0.61 & $-0.90$  \\[1ex]
			& & No Stars  &  &  &   \\[-1ex]
			\raisebox{1.5ex}{346} & \raisebox{1.5ex}{2.14$^{+0.1}_{-0.09}$}
			& No stars   &  &  &   \\[1ex]
			& 2.00   & 0.54 & ~ 0.19 & 0.93 & $-3.94$  \\[-1ex]
			\raisebox{1.5ex}{716} &
			0.68  & 0.11 & $-0.18$ & 0.40 & $-16.25$  \\[1ex]
			&2.07  & 0.57 & ~ 0.01 & 1.26 & $-11.22$  \\[-1ex]
			\raisebox{1.5ex}{1122} &
			1.73  & 0.34 & $-0.27$ & 0.99 & $-44.86$  \\[1ex]
			
			\botrule 
		\end{tabular}
		\label{table:DD-F:rotation}}
\end{table}
%%%%%%%%%%%%%%%%%%%%%%%%%%%%%%%%%%%%%%%%%%%%%%%%%%%%%%%%%%%%%%%%%%%%%
\begin{table}[ph]
	\caption{Properties of the rotating neutron star in the observed frequencies for DD-ME$\delta$ model} 
	{\begin{tabular}{@{}cccccc@{}} \hline\hline\\[1ex]
			$\nu$(Hz) & M(M$_{\odot}$)  & Z$_{p}$ & Z$_{eq}^{f}$ & Z$_{eq}^{b}$ & Q($10^{43}$gr.cm$^{2}$) \\ [1ex]
			\hline\\[1ex]
			& & No Stars   & & &    \\[-1ex]
			\raisebox{1.5ex}{25} & \raisebox{1.5ex}{2.01$\pm$0.04\cite{Antoniadis1233232}}
			& No Stars  & & &  \\[1ex]
			& & No Stars   &  & &   \\[-1ex]
			\raisebox{1.5ex}{317} & \raisebox{1.5ex}{1.97$\pm$0.04\cite{Demorest2010/10/28/print}}
			& 0.48  & ~ 0.33 & 0.64 &  $-0.69$  \\[1ex]
			& & No Stars   &  &  &   \\[-1ex]
			\raisebox{1.5ex}{346} & \raisebox{1.5ex}{2.14$^{+0.1}_{-0.09}$}
			& No stars   &  &  &   \\[1ex]
			& 1.97   & 0.53 & ~ 0.19 & 0.91 &   $-3.37$  \\[-1ex]
			\raisebox{1.5ex}{716} & 
			0.85  & 0.12 & $-0.19$ & 0.45&  $-11.93$  \\[1ex]
			& 2.03  & 0.56 & ~ 0.01 & 1.22 &  $-9.98$  \\[-1ex]
			\raisebox{1.5ex}{1122} & 
			1.67 & 0.33 & $-0.26$ & 0.97 &  $-30.90$  \\[1ex]
			
			\botrule
		\end{tabular}
		\label{table:DD-ME:rotation}}
\end{table}

%%%%%%%%%%%%%%%%%%%%%%%%%%%%%%%%%%%%%%%%%%%%%%%%%%%%%%%%
\begin{figure} 
	\includegraphics[width=8.5cm]{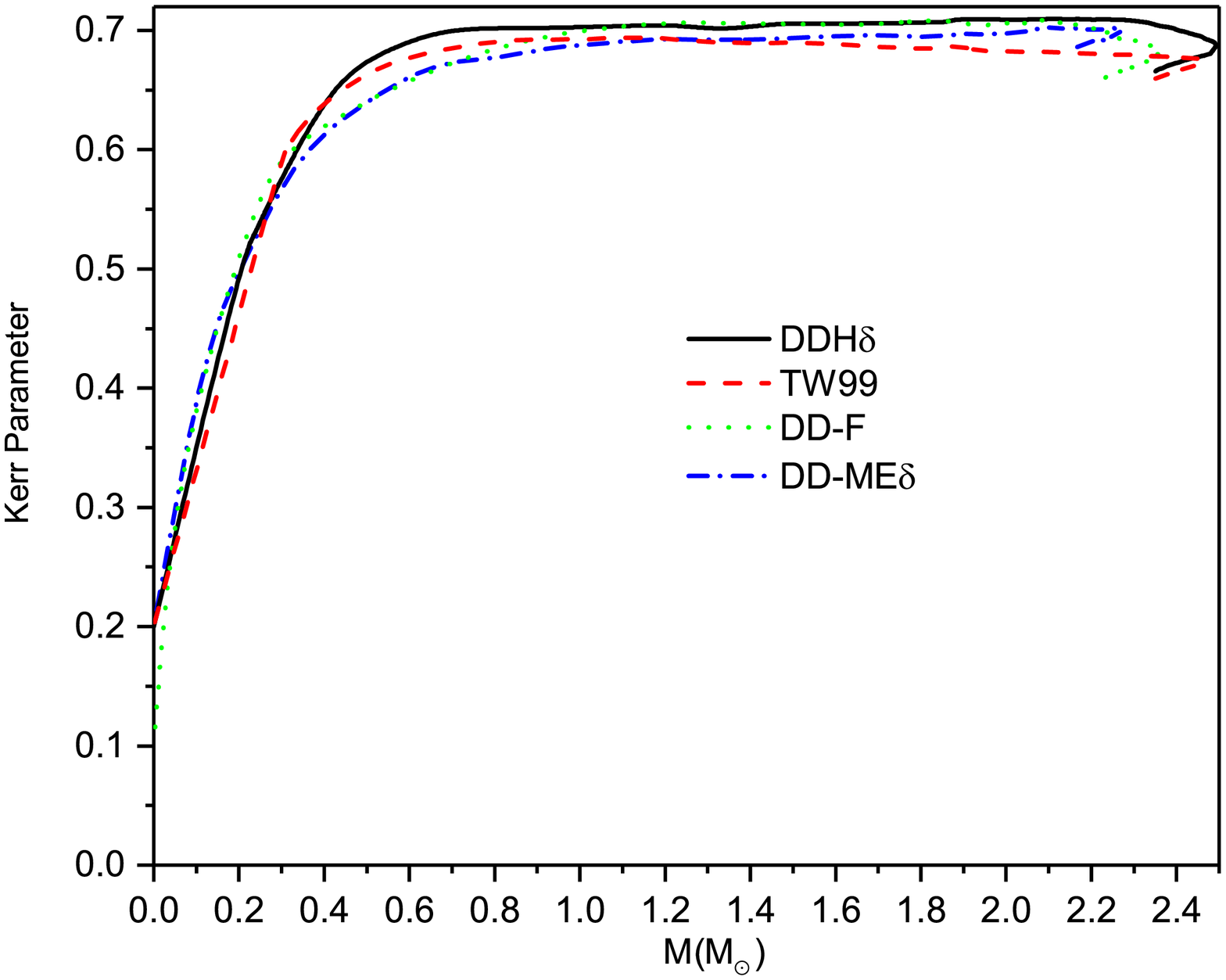}
\caption{(Color online) Kerr parameter, $a/M=cJ/(GM^{2})$ , as a function of the mass of neutron star along the Keplerian sequence. }\label{Kerr_Parameter}
\end{figure}

\subsection{Kerr Parameter}
The dimensionless angular momentum (Kerr parameter), $a/M=cJ/(GM^2)$ , is another important parameter can be used in the study of rotating neutron stars. Figure \ref{Kerr_Parameter} displays the Kerr parameter as funtion of the neutron star mass along the Keplerian sequence for the selected EOSs. It is obvious all graphs have the same maximum value, $(a/M)_{max}\approx 0.7$, approximately, similar to the results were obtained for the TM1, GM1 and NL3 EOSs\cite{PhysRevD.92.023007}. Such a maximum independent of the EOS implies the existence of universal values for the compactness and rotation-to-gravitational energy ratio of the neutron star. Because of this general behavior of Kerr parameter, it can be used to establish universal relations\cite{PhysRevLett.112.201102,10.1093/mnras/stw575,10.1093/mnrasl/slx178,PhysRevD.99.043004}. 

%%%%%%%%%%%%%%%%%%%%%%%%%%%%%%%%%%%%%%%%%%%%%%%%%%%%%%%%
%%%%%%%%%%%%%%%%%%%%%%%%%%%%%%%%%%%%%%%%%%%%%%%%%%%%%%%%
\subsection{Redshift} 
\begin{figure} 
	\includegraphics[width=8.5cm]{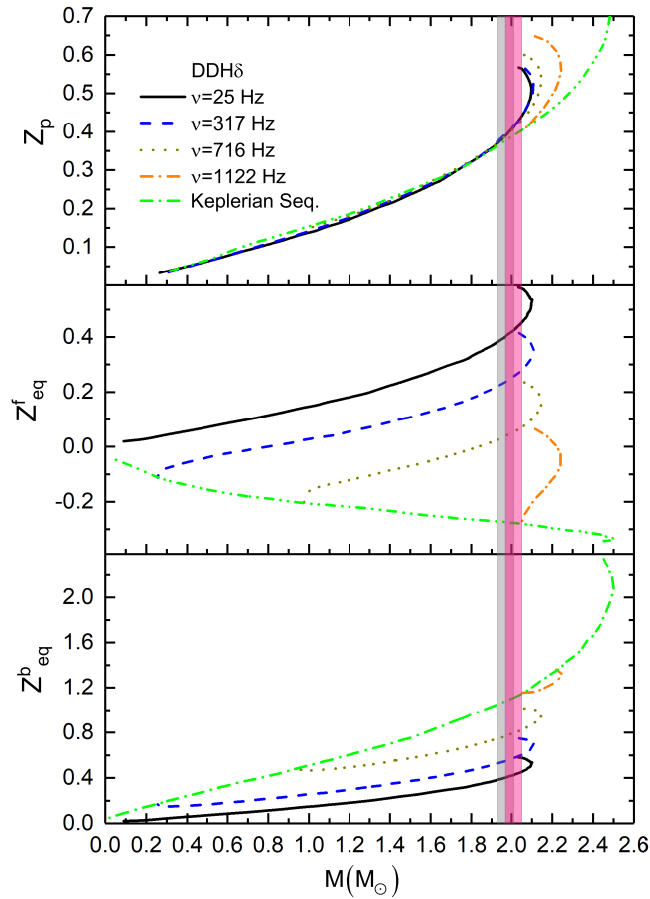}
\caption{(Color online) Polar (top panel), forward (middle panel) and backward (bottom panel) redshifts as a function of the mass of rotating neutron star for DDH$\delta$ model. The vertical bands show the masses of PSR J034+0432 (Pink) and PSR J1614-2230 (Grey). }\label{3P-DDHd_M_Red_Shift}
\end{figure}

\begin{figure} 
	\includegraphics[width=8.5cm]{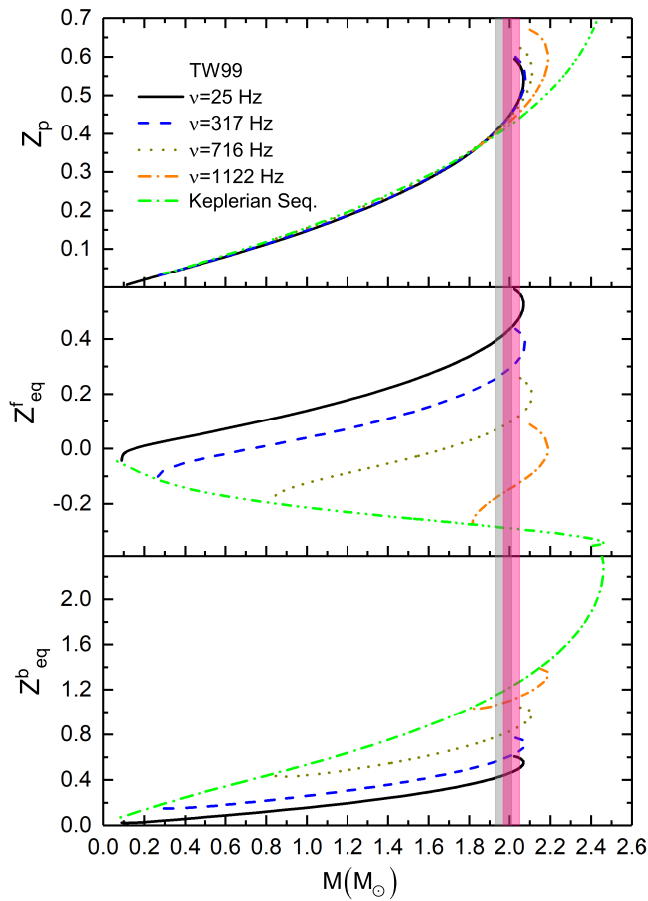}
	\caption{(Color online) Same as Figure \ref{3P-DDHd_M_Red_Shift}, but calculated for TW99 model.}\label{3P-TW99_M_Red_Shift}
\end{figure}

\begin{figure} 
	\includegraphics[width=8.5cm]{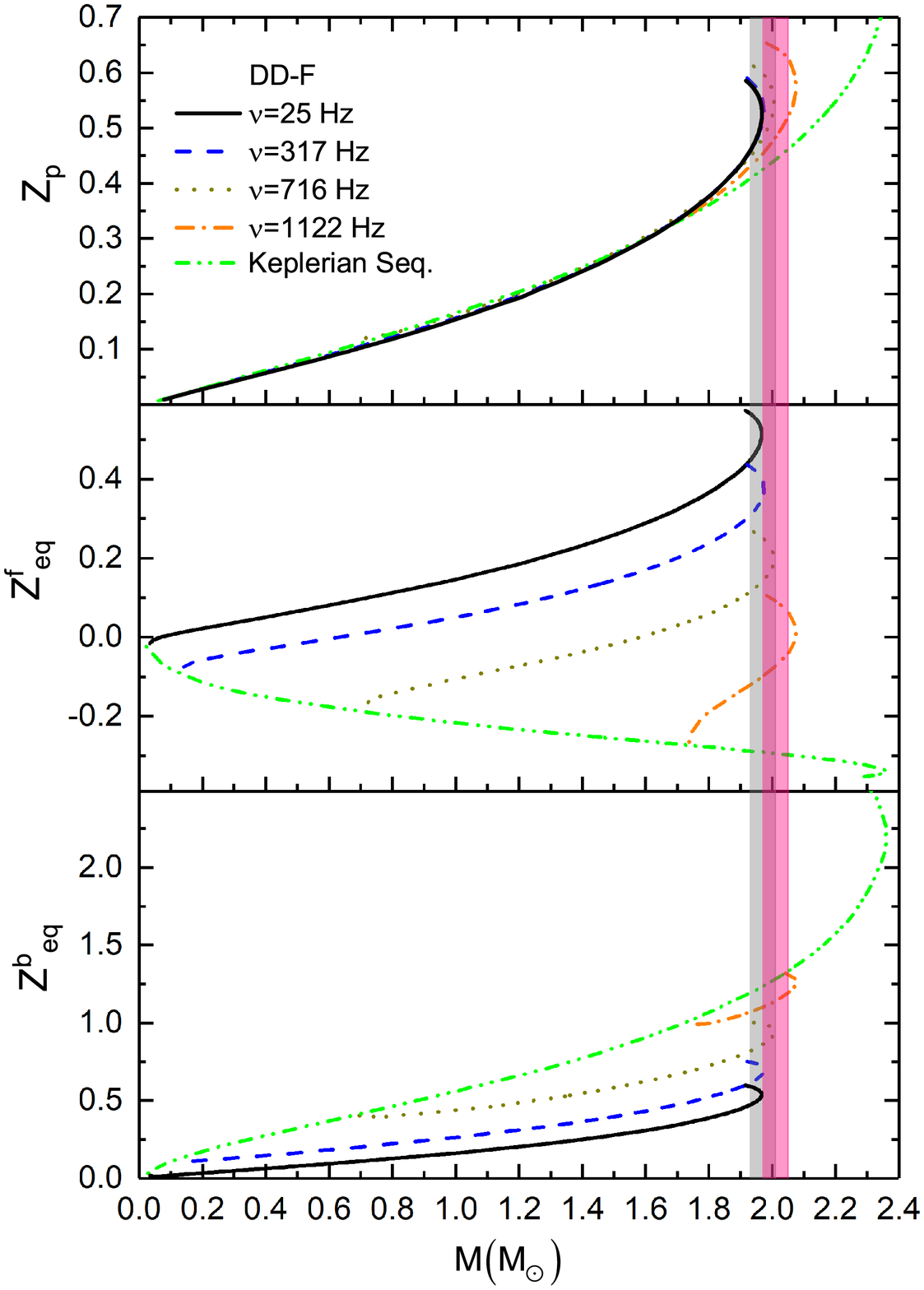}
	\caption{(Color online) Same as Figure \ref{3P-DDHd_M_Red_Shift}, but calculated for DD-F model.}\label{3P-DD-F_M_Red_Shift}
\end{figure}

\begin{figure} 
	\includegraphics[width=8.5cm]{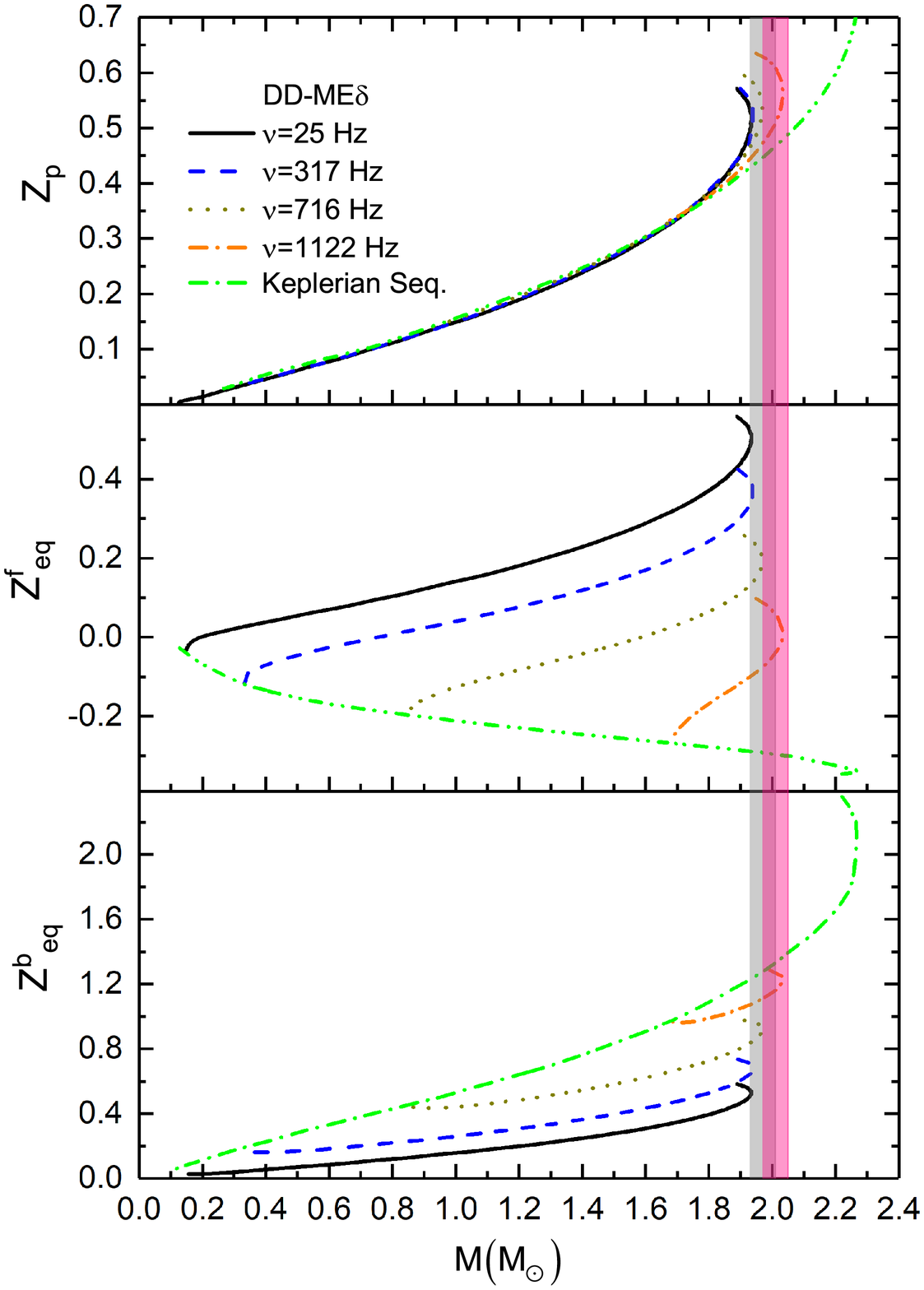}
\caption{(Color online) Same as Figure \ref{3P-DDHd_M_Red_Shift}, but calculated for DD-ME$\delta$ model.}\label{3P-DD-MEd_M_Red_Shift}
\end{figure}

Redshift, another key parameter is used for describing neutron stars\cite{1994ApJ...424..823C}. Redshift measurement can impose a constraint on mass-to-radius ratio, and in turn, the EOS of neutron stars. In the rotating star, if the radiation detector is directed in the polar plane of star, the polar redshift (gravitational redshift) can be measured. In contrast, if the radiation detector is tangentially directed to the star's surface, due to Doppler shift, two different redshifts, namely the forward redshift (blue shift) and the backward redshift can be measured. The polar redshift is defined as:
\begin{equation}
Z_{p}=e^{-2\nu_{p}}-1.
\end{equation}
The forward and backward redshifts are defined as follows\cite{NozawaT.1998}:
\begin{equation}
Z_{eq}^{f}=(\frac{1-v_{eq}}{1+v_{eq}})^{1/2}\frac{e^{-\nu_{eq}}}{1+R_{eq}\omega_{eq}
	e^{(\nu_{eq}-\psi_{eq})/2}}-1,
\end{equation}
\begin{equation}
Z_{eq}^{b}=(\frac{1+v_{eq}}{1-v_{eq}})^{1/2}\frac{e^{-\nu_{eq}}}{1-R_{eq}\omega_{eq}
	e^{(\nu_{eq}-\psi_{eq})/2}}-1,
\end{equation}
where the subscripts $p$ and $eq$ indicate values at the pole and the equatorial surface, respectively and $\nu$, $\omega$, and $\psi$ are the metric functions. For each selected frequency and EOS, the calculated redshifts are shown in Figures \ref{3P-DDHd_M_Red_Shift}-\ref{3P-DD-MEd_M_Red_Shift}. The range of redshifts is given in Tables \ref{table:DDHd:rotation}-\ref{table:DD-ME:rotation}. As shown in the top panels of these figures, the polar redshift is independent of the rotating frequency for M$<2.0 M_{\odot}$, and is equal to polar redshift in the Keplerian sequence. Moreover, the observed mass of low-frequency stars, J0740+6620, J0348+0432, and J1614-2230 rule out the soft EOS DD-F and DD-ME$\delta$. Therefore, the stiffer EOS such as TW99 and DDH$\delta$ are required to describe these stars.

\begin{figure} 
	\includegraphics[width=8.5cm]{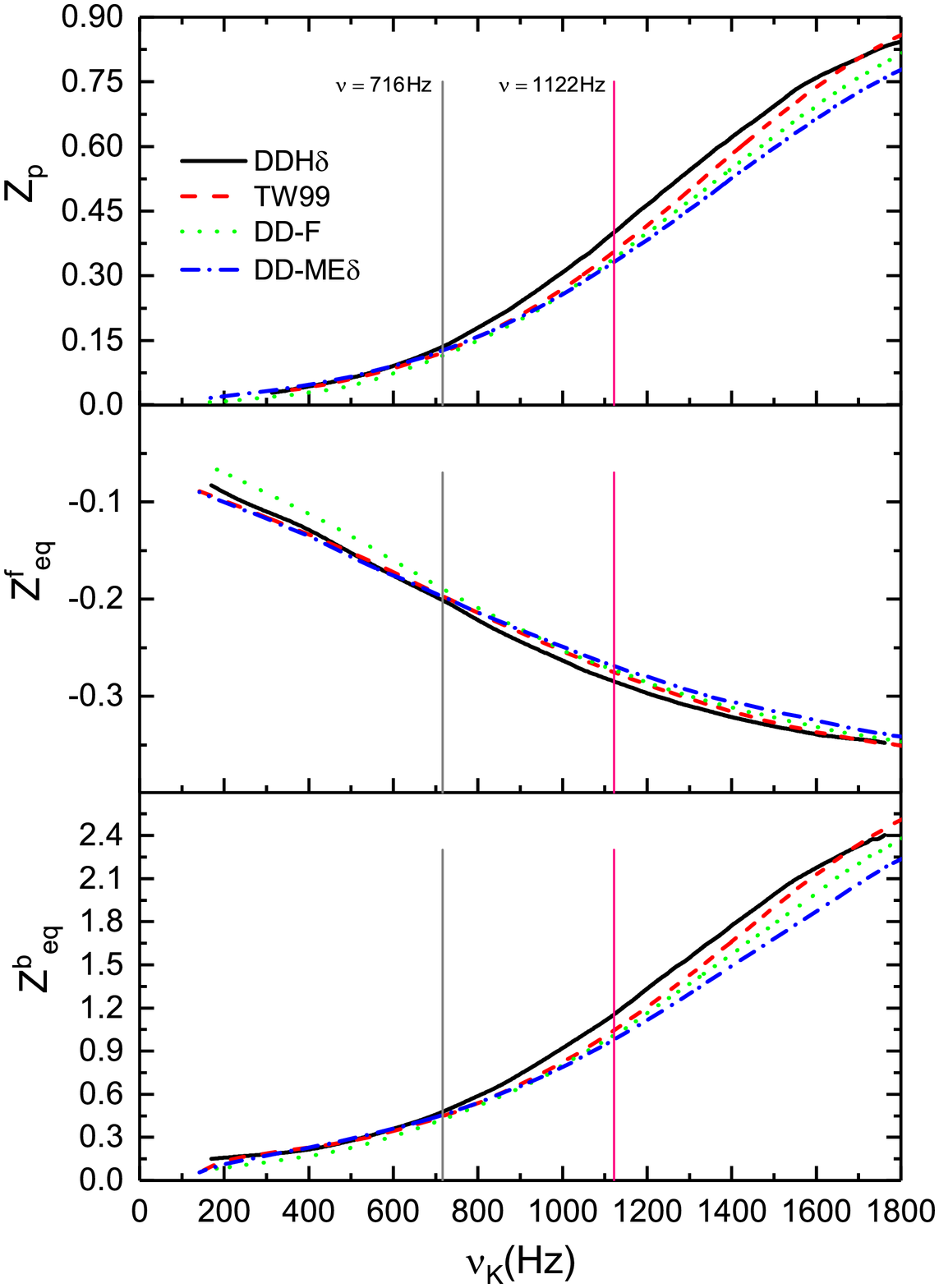}
	\caption{(Color online) Polar (top panel), forward (middle panel) and backward (bottom panel) redshift as a function of the Keplerian frequency. In this plot and hereafter, the vertical lines indicate the frequencies of the fastest rotating neutron stars, J1748-2446ad\cite{2006Sci...311.1901H} and J1739-285\cite{1538-4357-657-2-L97}. }\label{3P-f_Keplerian_Red_Shift}
\end{figure}

Figure \ref{3P-f_Keplerian_Red_Shift} displays the polar, forward and backward redshifts of neutron stars as a function of the Keplerian frequency.
One can conclude that if a neutron star rotates at a frequency equal to or greater than $\nu$= 716 Hz, its polar and backward redshifts are greater than 0.10 and 0.42, respectively, and its forward redshift is less than -0.20. Also, if it rotates at a frequency equal to or greater than $\nu$= 1122 Hz, its polar and backward redshifts are greater than 0.33 and 0.97, respectively, and its forward redshift is less than -0.28.

\begin{figure} 
	\includegraphics[width=8.5cm]{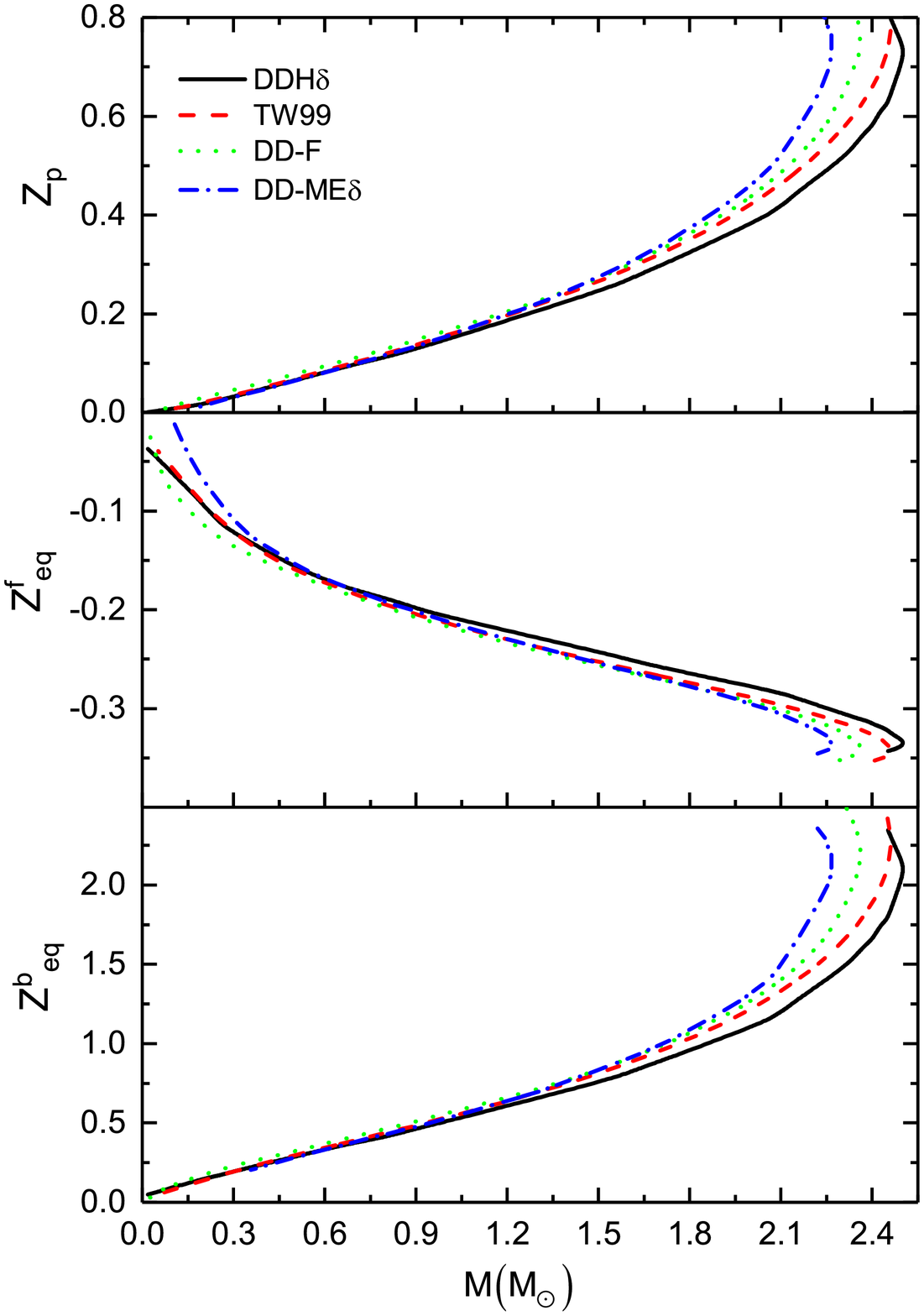}
\caption{(Color online) Polar (top panel), forward (middle panel) and backward (bottom panel) redshifts versus the mass of rotating star along the Keplerian sequence. }\label{3P-M_Keplerian_Red_Shift}
\end{figure}

Figure \ref{3P-M_Keplerian_Red_Shift} displays the redshifts of the rotating star as a function of the star mass along the Keplerian sequence. As clearly seen, the extremum values of redshifts reach $Z_{p}\approx$ 0.8, $Z^{f}_{eq}\approx$ -0.3 and $Z^{b}_{eq}\approx$ 2.2 for all the selected EOSs. According to the last two figures, the redshifts along the Keplerian sequence show a general behavior independent of EOS, approximately. Hence, this behavior should be evaluated by the other EOSs and if this is confirmed, universal relations can be established for these parameters\cite{PhysRevLett.112.201102,10.1093/mnras/stw575,10.1093/mnrasl/slx178,PhysRevD.99.043004}.

The neutron star redshift database provided by measuring $\gamma$-ray burst redshifted annihilation lines has been interpreted as gravitationally redshifted 511 kev e$^{\pm}$ pair annihilation from the surface of neutron star\cite{liang1986gamma}. If this interpretation is correct, it will support a neutron star redshift range of $0.2\le Z_P \le 0.5$, with the highest concentration in the range of $0.25\le Z_P \le 0.35$. Our results show the gravitational mass of this neutron star is in the range of [1.5, 1.9]M$_{\odot}$. On the other hand, Figure \ref{3P-f_Keplerian_Red_Shift} shows the rotating frequency of this star is in the range of [950, 1050]Hz if it rotates at its Keplerian frequency. 

The two absorption lines in the spectrum of the young isolated neutron star, 1E1207.4-5209, imply a redshift of $0.12 \le Z_{P} \le 0.23$\cite{2002ApJ...574..61}. As seen in Figure \ref{3P-DDHd_M_Red_Shift}-\ref{3P-DD-MEd_M_Red_Shift}, the gravitational mass of this star is in the range of $[0.95,1.5]M_{\odot}$. Figure \ref{3P-f_Keplerian_Red_Shift} shows if it rotates at the Keplerian frequency, its frequency is in the range of [700, 850]Hz. 
Two different mass ranges of, $2.00^{+0.07}_{-0.24}M_{\odot}$ and $1.5^{+0.4}_{-1.0}M_{\odot}$\cite{10.1093/mnras/stx1452} are obtained for the low-mass X-ray binary neutron star, EXO 0748-676, with rotating frequency of 552 Hz\cite{Galloway_2010}, a polar redshift of $Z_P\approx 0.35$\cite{Cottam2002}. If an approximate mass  $2.00~M_{\odot}$ is confirmed, none of the considered EOSs in this study is appropriate for describing this star, and stiffer EOSs should be used for this purpose. On the other hand, if an approximate mass $1.5M_{\odot}$ is confirmed, the calculated redshift of $Z_{P}\approx 0.3$, will be close to the measured value, and these EOSs are appropriate for describing this star. According to Figure \ref{3P-f_Keplerian_Red_Shift}, the approximate Keplerian frequency equals 1000 Hz, and the star is rotating away from the disintegration threshold.
The predicted mass of the isolated 205.33 Hz millisecond pulsar PSR J0030+0451 was obtained from the analysis of the NICER data of the thermal X-ray wave from this object\cite{Miller_2019}. The reported mass is in the range of $1.44^{+0.15}_{-0.14}M_{\odot}$. The top panels of Figures \ref{3P-DDHd_M_Red_Shift}-\ref{3P-DD-MEd_M_Red_Shift} show its polar redshift can be in the range of $0.2 \le Z_{P} \le 0.3$.

%%%%%%%%%%%%%%%%%%%%%%%%%%%%%%%%%%%%%%%%%%%%%%%%%%%%%%%%
%%%%%%%%%%%%%%%%%%%%%%%%%%%%%%%%%%%%%%%%%%%%%%%%%%%%%%%%
\subsection{Quadrupole Moment}      
\begin{figure} 
	\includegraphics[width=8.5cm]{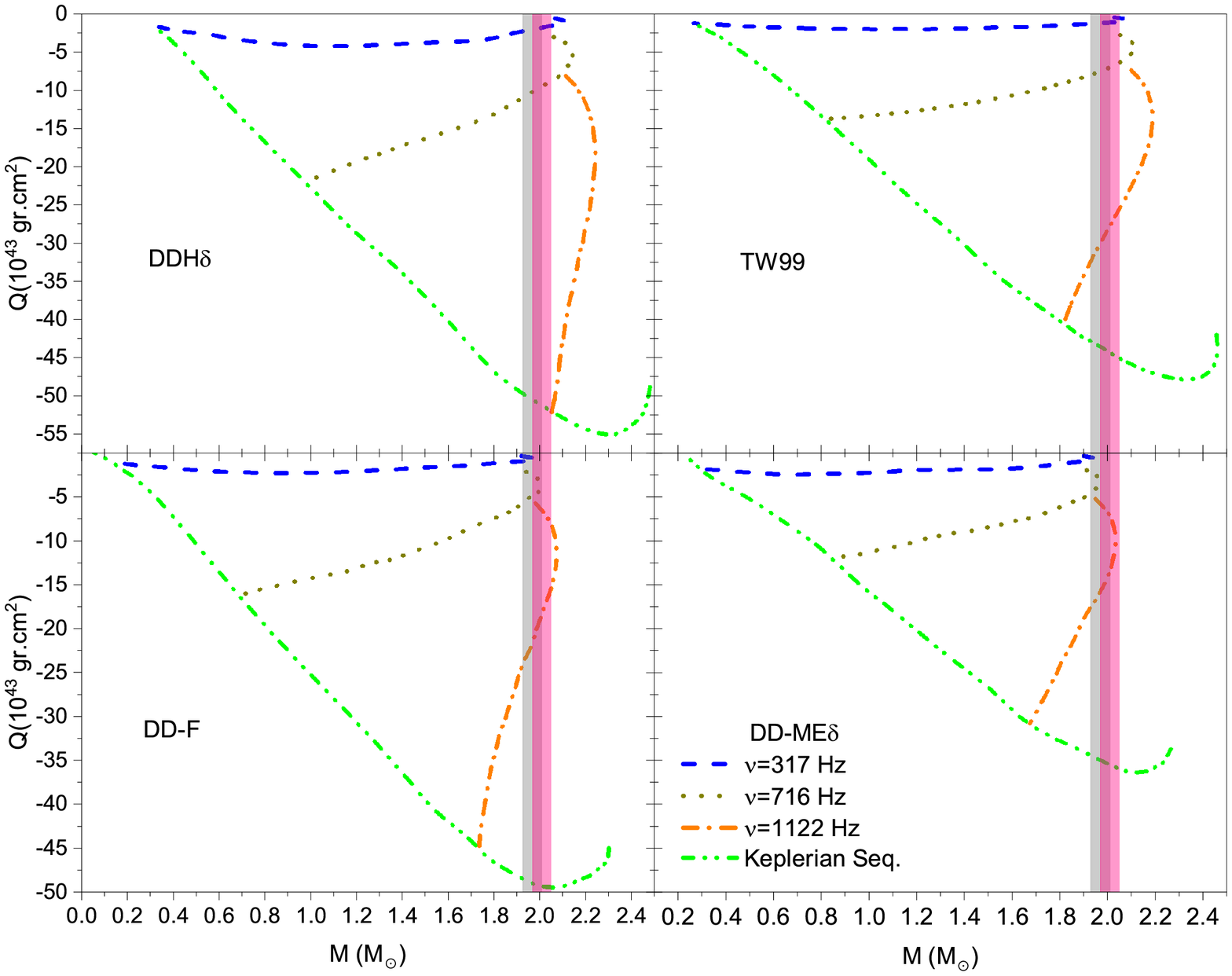}
	\caption{(Color online) The quadrupole moment at the rotating frequencies of 317, 716, and 1122 Hz and along the Keplerian sequence versus the mass of the star.  }\label{4P-M_Quad}
\end{figure}

\begin{figure} 
	\includegraphics[width=8.5cm]{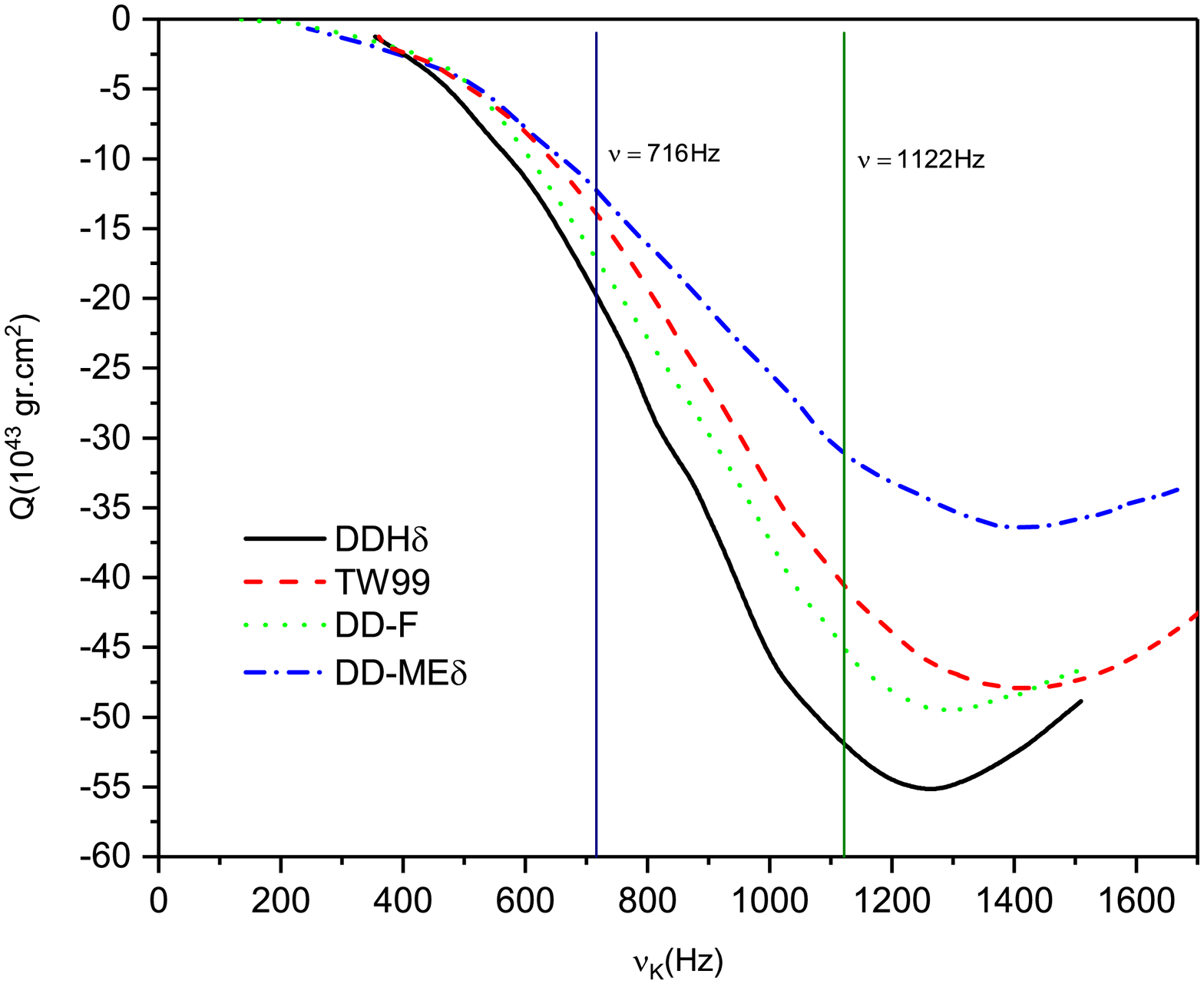}
\caption{(Color online) The quadrupole moment as a function of the Keplerian frequency.}\label{f-Quad_Keplerian}
\end{figure}

A fast-rotating neutron star is, in fact, a deformed distribution of stellar mass. This elliptical object causes a distortion in the external gravitational field, which is measured by the quadrupole moment Q. Despite its small-magnitude effect, it is comparable to that caused by the general relativistic spin-spin interaction\cite{PhysRevD.47.R4183,PhysRevD.49.6274}. The quadrupole moment has been discussed in literature\cite{1977MNRAS.179..483M,PhysRevD.52.5707,0004-637X-512-1-282,1994A&A...291..155S}. The following definition has been presented for the quadrupole moment:
\begin{equation}\label{tildeQ}
\tilde{Q}=-\frac{3}{8\pi}\int\sigma_{LnN}(cos^{2}\theta-\frac{1}{3})r^{4}sin\theta drd\theta d\phi.
\end{equation}
A systematic error occurs in this calculation. It is assumed that the metric expressed in the used coordinate system has an asymptotic behavior similar to the Schwarzschild metric behavior up to some order. Pappas and Apostolatos \cite{PhysRevLett.108.231104} corrected the quadrupole moment of the rotating neutron star and defined it as follows:
\begin{equation}
Q=\tilde{Q}-\frac{4}{3}(\frac{1}{4}+b)M^{3},
\end{equation}
where $\tilde{Q}$ is defined in Eq. (\ref{tildeQ}) and b is defined by Eq. (3.37) of Ref. \cite{Rotating..Relativistic..Stars..2013} and M is the gravitational mass of the star. The results of calculations for each observed star is given in Tables \ref{table:DDHd:rotation}-\ref{table:DD-ME:rotation} and Figure \ref{4P-M_Quad}. A quadrupole moment less than zero, Q$<$0, indicates an oblate spheroid. Figure \ref{f-Quad_Keplerian} displays the quadrupole moment of neutron star as a function of the Keplerian frequency. It can be deduced that if the neutron star rotates at a frequency equal to or greater than $\nu$= 716 Hz, its quadrupole moment is equal to or less than -12.27$\times 10^{43}~gr.cm^2$. If it rotates at a frequency equal to or greater than $\nu$= 1122 Hz, its quadrupole moment is equal to or less than -31.01$\times 10^{43}~gr.cm^2$ or less.   

%%%%%%%%%%%%%%%%%%%%%%%%%%%%%%%%%%%%%%%%%%%%%%%%%%%%%%%%

%%%%%%%%%%%%%%%%%%%%%%%%%%%%%%%%%%%%%%%%%%%%%%%%%%%%%%%%
%%%%%%%%%%%%%%%%%%%%%%%%%%%%%%%%%%%%%%%%%%%%%%%%%%%%%%%%
%%%%%%%%%%%%%%%%%%%%%%%%%%%%%%%%%%%%%%%%%%%%%%%%%%%%%%%%
\section{Conclusions}
Rotating neutron stars were investigated by the density-dependent mean-field interactions using the DDH$\delta$, TW99, DD-F and DD-ME$\delta$ EOSs. The stability region and the equilibrium sequences were constructed at the observed frequencies of 25, 317, 346, 716 and 1122 Hz. The most massive neutron stars yet found can be described by the DDH$\delta$ and TW99 EOSs. On the other hand, DD-F and DD-ME$\delta$ models were so soft to sustain the maximum observed masses at the rotating frequencies of 25, 317, and 346 Hz. Considering the intersection of the secular axisymmetric instability with the Keplerian sequence, one can conclude the rotation frequency of the fastest observed star, 1122 Hz, does not impose a strong constraint on the EOS.  

The minimum and maximum masses of the fastest rotating stars were obtained within the bounds M$_{min}$=[0.68, 0.98] M$_{\odot}$ and M$_{max}$=[1.97, 2.14] M$_{\odot}$ at the rotation frequency of $\nu$= 716 Hz and M$_{min}$=[1.67, 2.04] M$_{\odot}$ and M$_{max}$=[2.03, 2.24] M$_{\odot}$ at the rotating frequency of $\nu$= 1122 Hz.

The Kerr parameter reached a maximum value of $(a/M)_{max}\approx$ 0.7. Given the Kerr parameter of black hole, a$/$M= 1, by definition, the gravitational collapse of a rotating neutron star cannot lead to the Kerr black hole.

The polar, forward and backward redshifts were evaluated in all selected rotation frequencies and the Keplerian sequence. Interestingly, each redshift reached extremums of $Z_{p}\approx$ 0.8, $Z^{f}_{eq}\approx$ -0.3 and $Z^{b}_{eq}\approx$ 2.2 independent of the EOS in the Keplerian sequence.

Comparison of the results of this study with the measured polar redshift indicates that:

1) Our results are consistent with that announced by Liang \textit{et al.}, and the selected EOS properly described the star considered in their study. Our results show the gravitational mass of this star may be in the range of [1.5, 1.9]$M_{\odot}$. Also, if the star rotates at its Keplerian frequency, this frequency may be in the range of [950, 1050]Hz. 

2) Considering the redshift of the isolated neutron star 1E1207.4+5200, its gravitational mass and Keplerian frequency may be in the range of [0.95, 1.5]$M_{\odot}$ and [700, 800]Hz, respectively.

3) Confirming a mass of $2.00^{+0.07}_{-0.24}M_{\odot}$ for EXO 0748-676, none of the considered EOSs seems appropriate to describe this star. On the other hand, if a mass of $1.50^{+0.4}_{-1.0}M_{\odot}$ is confirmed, our result, $Z_{P}\approx 0.3$, will be close to the measured value, and these EOSs are able to properly describe this star. The approximate Keplerian frequency of this star equals 1000Hz, and it is probably rotating away from the disintegration threshold.

{4) considering the reported mass of the isolated millisecond pulsar PSR J0030+0451, its polar redshift may be in the range of $0.2 \le Z_{P} \le 0.3$.
	
	All calculated quantities including the gravitational mass, redshifts and quadrupole moment along the Keplerian sequence were plotted versus the Keplerian frequency. According to the diagrams, only neutron stars with approximate mass, polar and backward redshifts greater than 1.7 M$_{\odot}$, 0.33 and 0.97, respectively, and with forward redshift, and quadrupole moment smaller than -0.28, and $-31.01\times 10^{43}gr.cm^{2}$, respectively, can rotate at a frequency equal to or greater than $\nu$= 1122 Hz.
	The unmeasured quantities redshifts and quadrupole moment that may be measured in the future were calculated. Given the general behavior of the redshifts and Kerr parameter, universal relations can be constructed for these quantities.  
	%%%%%%%%%%%%%%%%%%%%%%%%%%%%%%%%%%%%%%%%%%%%%%%%%%%%%%%%
	%%%%%%%%%%%%%%%%%%%%%%%%%%%%%%%%%%%%%%%%%%%%%%%%%%%%%%%%
	\section{Acknowledgments}
	R. Riahi thanks Jorge A. Rueda from ICRANet in Pescara for the valuable discussions.
	%%%%%%%%%%%%%%%%%%%%%%%%%%%%%%%%%%%%%%%%%%%%%%%%%%%%%%%%
	%%%%%%%%%%%%%%%%%%%%%%%%%%%%%%%%%%%%%%%%%%%%%%%%%%%%%%%%
	\clearpage
	%\bibliographystyle{apsrev}
	%\bibliography{density-dependent}

\end{document}